Collisions of ultracold $^{23}$Na$^{87}$Rb molecules with controlled chemical reactivities

X. Ye,[1] M. Guo,[1] M. L. González-Martínez,[2] G. Quéméner,[2] D. Wang[1,3]*

[1]Department of Physics, The Chinese University of Hong Kong, Hong Kong SAR, China.
[2]Laboratoire Aimé Cotton, CNRS, Université Paris-Sud, ENS Paris-Saclay, Université Paris-Saclay, 91405 Orsay Cedex, France.
[3]The Chinese University of Hong Kong Shenzhen Research Institute, Shenzhen, China.

*To whom correspondence should be addressed; E-mail: djwang@cuhk.edu.hk.

**Abstract**
The collision of molecules at ultracold temperatures is of great importance for understanding the chemical interactions at the quantum regime. While much theoretical work has been devoted to this, experimental data are only available sparsely mainly due to the difficulty in producing ground-state molecules at ultracold temperatures. We report here the creation of optically trapped samples of ground-state bosonic sodium-rubidium molecules with precisely controlled internal states and, enabled by this, a detailed study on the inelastic loss with and without the NaRb + NaRb → Na$_2$ + Rb$_2$ chemical reaction. Contrary to intuitive expectations, we observed very similar loss and heating, regardless of the chemical reactivities. In addition, as evidenced by the reducing loss rate constants with increasing temperatures, we found that these collisions are already outside the Wigner region even though the sample temperatures are sub-microkelvin. Our measurement agrees semi-quantitatively with models based on long-range interactions, but calls for a deeper understanding on the short-range physics for a more complete interpretation.

**MAIN TEXT**

**INTRODUCTION**
Collisional study is a primary way for understanding the interaction between particles, for example, much of today's knowledge on chemical reaction dynamics owes to pioneering studies on elementary reaction processes with the crossed molecular beam experiments (*1*). In recent years, thanks to the development of ultracold technologies, controlling the external motion between particles to the unprecedented single and the lowest relative angular momentum level has become routinely available. In this quantum regime, theoretical studies have suggested early on that chemical reactions can happen efficiently with the help of quantum effects such as tunneling (*2*). In 2010, such ultracold chemical reaction was first confirmed via measuring the loss in a gaseous sample of chemical reactive $^{40}$K$^{87}$Rb molecules (*3*).

Besides its relevance in ultracold chemistry, understanding the basic molecular interaction at ultra-low collision energies is also a necessary step toward unveiling the full potential of ultracold polar molecules (UPMs) on a very broad range of applications in quantum simulation of exotic many-body physics (*4*, *5*), quantum information (*6*, *7*) and precision measurements for fundamental constants. In fact, in several recently created UPMs, including fermionic $^{23}$Na$^{40}$K (*8*) and bosonic $^{87}$Rb$^{133}$Cs (*9*, *10*) and $^{23}$Na$^{87}$Rb (*11*) which are all non-reactive and thus thought to be immune to inelastic collisions, large losses were still observed. Presently, this unexpected loss has become the major obstacle for advancing the field since it prohibits the quest for quantum degenerate UPMs via evaporative cooling. Although there is still no definitive explanation for this loss, it is generally believed that the four-atom collision complex formed following the encounter of two molecules is playing a key role (*12*).

In this work, we present a direct comparison on the inelastic losses of UPMs with and without the bimolecular chemical reaction energetically allowed. This study is performed with optically trapped ultracold gases of bosonic $^{23}$Na$^{87}$Rb molecules with their chemical reactivity controlled via vibrational state manipulation (hereafter, $^{23}$Na and $^{87}$Rb will be denoted as Na and Rb, respectively). Vibrational excitation has already been established as an efficient way for controlling both the chemical reaction rate and the outcome since the 1970s (*13–16*), but it has never been exploited in ultracold molecular samples.

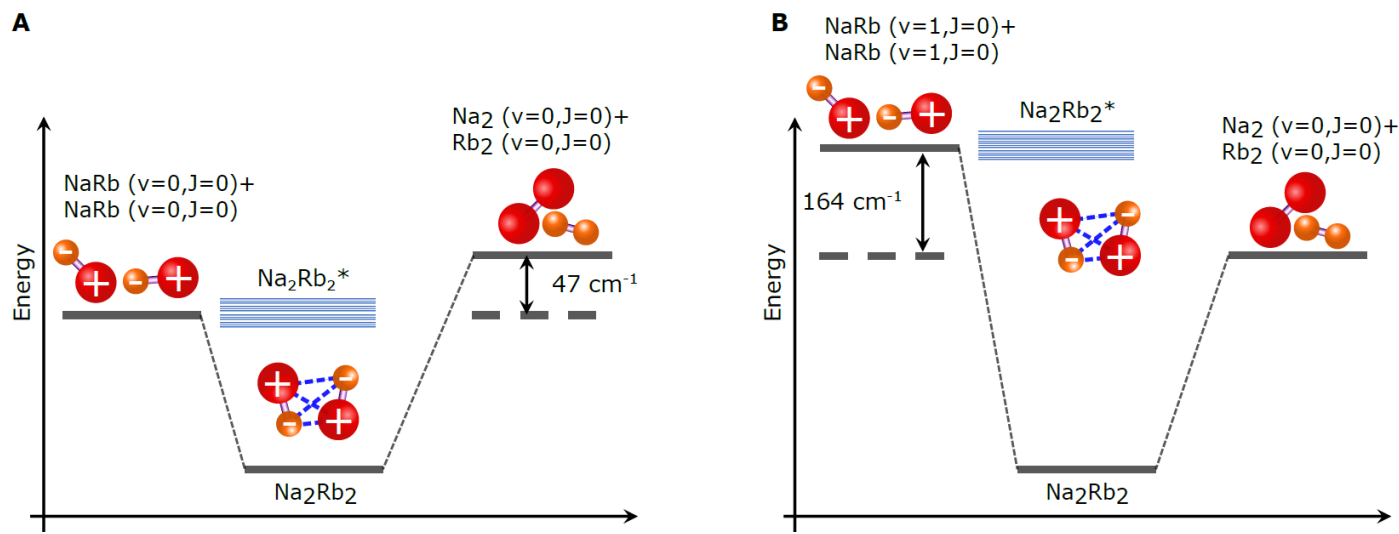

**Fig. 1. Control the chemical reactivity of NaRb molecules by vibrational excitation.** As shown are the schematic reaction coordinates for the NaRb + NaRb → Na$_2$ + Rb$_2$ process. (**A**) In the lowest rovibrational level ($v = 0$, $J = 0$), the reaction is endothermic by 47 cm$^{-1}$ and thus forbidden at ultracold temperatures. (**B**) In the first excited rovibrational level ($v = 1$, $J = 0$), the same reaction is already exothermic by 164 cm$^{-1}$ and thus allowed. Molecules can also relax from $v = 1$ to $v = 0$ following the collision, but experimentally this cannot be distinguished from chemical reactions. The ground Na$_2$Rb$_2$ tetramer level, which has much lower energy than both the reactant and product molecule pairs, is also shown. Near the NaRb + NaRb collision threshold the density of Na$_2$Rb$_2$* states is estimated to be too large to be resolved. As a result, the collision is in the highly-resonant regime (*12*).

The main feature of the Na and Rb diatomic molecules which enables our investigation are illustrated schematically in Fig. 1. For NaRb molecules in the absolute ground state, the reaction

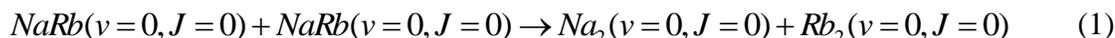
$$NaRb(v=0, J=0) + NaRb(v=0, J=0) \rightarrow Na_2(v=0, J=0) + Rb_2(v=0, J=0) \quad (1)$$

is endothermic by 47 cm$^{-1}$ (*17–20*). Here $v$ and $J$ are the vibrational and rotational quantum numbers, respectively. The situation is starkly different for NaRb in the first excited vibrational level since the reaction

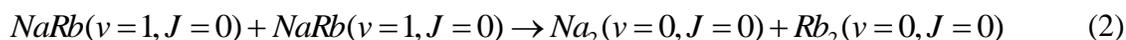
$$NaRb(v=1, J=0) + NaRb(v=1, J=0) \rightarrow Na_2(v=0, J=0) + Rb_2(v=0, J=0) \quad (2)$$

is exothermic by 164 cm$^{-1}$ (17–20); thus energetically, the Na$_2$ and Rb$_2$ product molecules in many rotational levels and partial waves can be produced. This reaction, as well as all other bimolecular reactions between alkali molecules (17, 21), should be barrierless with the Na$_2$Rb$_2$ potential well lying thousands of wavenumbers below both the NaRb + NaRb and the Na$_2$ + Rb$_2$ asymptotic limits.

Importantly, the long-range bimolecular NaRb interaction coefficient for ($v = 1, J = 0$) molecules ($C_6 = 1.533 \times 10^6$ a.u.) is essentially identical to that for ($v = 0, J = 0$) molecules ($C_6 = 1.525 \times 10^6$ a.u.) (22–24); thus, comparing collisions of Eqs. 1 and 2 should reveal directly the difference in their short-rang physics. We find that, somewhat surprisingly, the losses have a very weak dependence on the chemical reactivity. Our result can be explained semi-quantitatively by the model based on complex formation, but it also points out the importance of further understanding the properties and the post-formation dynamics of the complex.

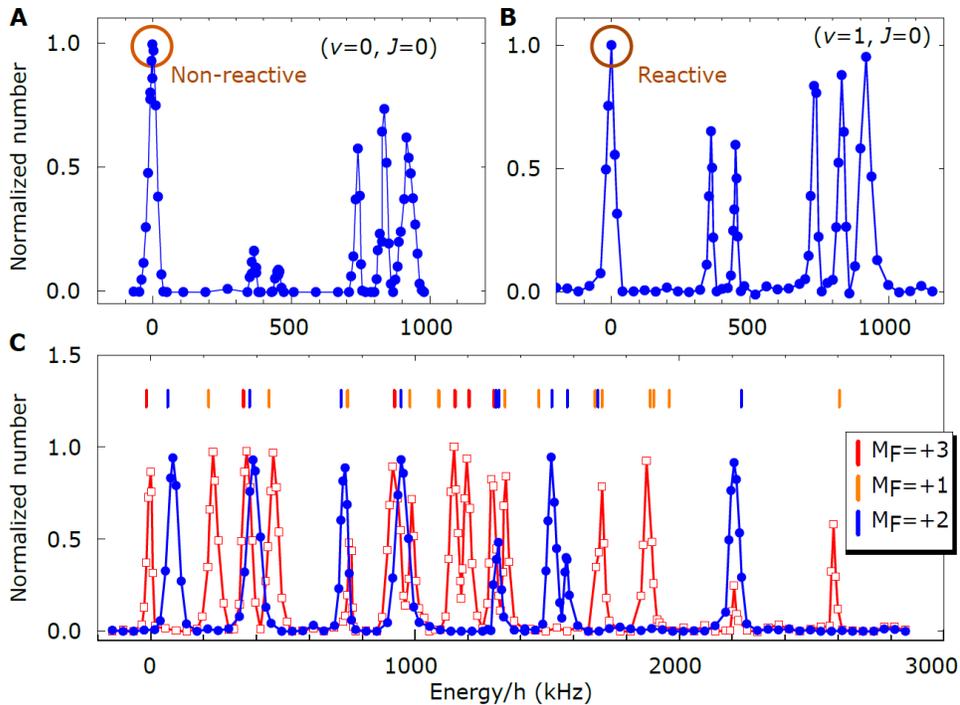

**Fig. 2. High resolution internal state control.** (**A**), (**B**) and (**C**) show well-resolved hyperfine structures of the rovibrational ground state ($v = 0, J = 0$), the first excited rovibrational state ($v = 1, J = 0$) and the rotationally excited ($v = 0, J = 2$) states, respectively. The two $M_F = 3$ hyperfine levels marked by the open circles in (**A**) and (**B**) are used in this work. (**C**) contains spectra obtained with different Raman laser polarization combinations for demonstrating $M_F$ control. The color-coded vertical bars in (**C**) mark the predicted hyperfine line positions.

## RESULTS
### Loss measurements with different chemical reactivities
The starting point of our experiment is an optically trapped sample of weakly-bound NaRb Feshbach molecules (FMs) (25). To study and compare the collisions in Eqs. 1 and 2, we transfer the FMs to the target levels directly by a stimulated Raman adiabatic passage (STIRAP) (11). With carefully chosen Raman laser frequency, power, and polarization combinations, we can create high density samples of NaRb molecules in ($v = 0, J = 0$) and ($v = 1, J = 0$) with the nuclear hyperfine structures fully resolved (Fig. 2A and B). We can also prepare molecules in the ($v = 0$, $J = 2$) level which have even richer hyperfine Zeeman structures due to the nuclear spin-rotation coupling (Fig. 2C). However, investigating collisions with rotational relaxation is beyond the scope of this work.

Since we rely on absorption imaging of atoms for detection, a reversed STIRAP has to be applied to transfer the ground-state molecules back to the Feshbach state to be dissociated into atoms (*11*). By varying the delay time between the two STIRAP sequences, the evolution of the ground-state molecule number $N$ can be measured. To increase the detection sensitivity, the imaging system is set up along the long axis of the sample for higher integrated optical depths. This has allowed us to take long duration measurements to cover a large range of molecule numbers. Fig. 3A shows example measurements for both ($v = 0$, $J = 0$) and ($v = 1$, $J = 0$) molecules with initial to final number ratios of nearly 40. Accompanying the fast losses, rapid temperature increases are also observed (Fig. 3B) due to the preferential removal of lower energy molecules characteristic of all inelastic collisions (*26*, *27*).

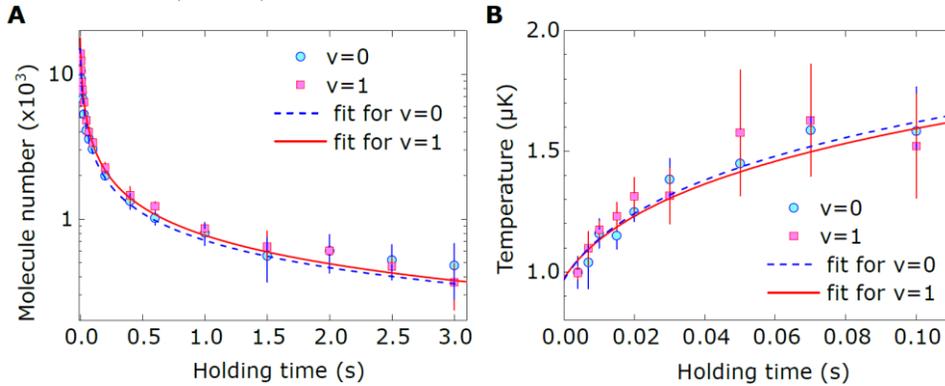

**Fig. 3. Inelastic collisions with different chemical reactivities.** (**A**) and (**B**) show time evolutions of molecule numbers and temperatures, respectively, for both non-reactive ($v = 0$, $J = 0$) (filled circles) and reactive ($v = 1$, $J = 0$) (filled squares) samples. The temperatures measurement, which stops at 0.1 second due to reduced SNR following the time-of-flight expansion, is obtained separately from the number evolution with samples of essentially identical conditions. Error bars represent one standard deviation. The blue dashed and red solid curves are fitting results using Eq. 3 with temperature-dependent loss rate constants obtained from Fig. 4 (see text for details). The measured trap oscillation frequencies are [$\omega_x$, $\omega_y$, $\omega_z$] = $2\pi \times$ [217(3),208(3),38(2)] Hz for the ($v = 0$, $J = 0$) molecules and $2\pi \times$ [219(3), 205(2), 40(2)] Hz for the ($v = 1$, $J = 0$) molecules, respectively. The calculated initial peak densities can reach $6 \times 10^{11}$ cm$^{-3}$.

A very striking feature in Fig. 3 is the number loss and the heating of the two cases are nearly identical, despite their distinctly different chemical reactivities and the very large range of number variations. To obtain a more quantitative picture, we fit simultaneously the number and temperature evolutions for each case to the two-body inelastic loss model (*26*)

$$\frac{dN(t)}{dt} = -\beta A \frac{N(t)^2}{T(t)^{3/2}}$$
$$\frac{dT(t)}{dt} = \beta A N(t) \frac{1/4 + h_0}{T(t)^{1/2}} \quad (3)$$

to extract the loss rate constant $\beta$. Here, $A = (\bar{\omega} m/4\pi k_B)^{3/2}$ is a constant with $m$ the mass of the molecule, and $\bar{\omega}$ is the geometric mean of the trap frequencies which is measured to be $2\pi \times 119.9(1.5)$ Hz for $(v = 0, J = 0)$ molecules and $2\pi \times 122.0(1.5)$ Hz for $(v = 1, J = 0)$ molecules. The $h_0$ term takes care of the momentum dependence of $\beta$ as well as any other heating contributions. We note that although the number loss and heating in Fig. 3 appear identical for the two cases, since the trap frequencies are not the same, the $\beta$ values are actually different.

**Temperature dependence of loss rate constants**
Initially, we used a constant $\beta$ to fit the whole data set only to find that the values obtained vary with the initial sample temperatures. This indicates that $\beta$ is changing during the course of the collisions following the temperature variation; thus, the collisions are not in the Wigner threshold region (*28*, *29*) in which $\beta \propto T^l$ should be a constant for s-wave ($l = 0$) collisions. This non-Wigner behavior can be understood by the much smaller characteristic temperature of the van der Waals potential $T_{vdW} = (\hbar^2/mr_6^2)(1/k_B)$ for NaRb (*30*), which is only 2.8 µK due to the very large long-range $C_6$ coefficients (*22–24*). Here $\hbar$ is the Planck constant $h$ over $2\pi$, $k_B$ is the Boltzmann's constant, and $r_6 = (mC_6/\hbar^2)^{1/4}$ is the characteristic length. The Wigner threshold region (*28*, *29*), which requires $T \ll T_{vdW}$, is thus not reached with the current sample temperature.

To obtain $\beta$ without knowing its exact $T$ dependence, we have used the following procedure (*26*). We first divide a full measurement into several segments. During each segment the temperature increases by less than 20% and thus $\beta$ should change very little. We then fit each segment to Eq. 3 with the initial number, the initial temperature, the additional heating term $h_0$, and a constant $\beta$ as free parameters. The same procedure is repeated for several data sets with initial temperatures ranging from 370 nK to 1.4 µK. Following the temperature increase, this has allowed us to sample $\beta$ from 390 nK to 1.85 µK for the non-reactive molecules and 500 nK to 1.85 µK for the reactive molecules. Fig. 4 shows the results with the mean temperature of each corresponding segment as the horizontal axis. Within the temperature range covered, the lowest and the highest $\beta$ values are only different by a factor of two. This small dynamical range and the relatively large error bars prevented us from confirming any structures in $\beta$, although an overall decrease toward higher temperatures is obvious for both cases. In addition, the measured $\beta$ for non-reactive samples are all larger than those of the reactive ones within the same temperature range. In the log-log scale, forcing linear fits to the data points in Fig. 4, slopes of $b = -0.38(4)$ and $b = -0.27(8)$ can be obtained for non-reactive and reactive samples, respectively. Thus approximately, during each set of measurements, the loss rate constant will follow a power law function $\beta(T) = \beta_0(T/T_0)^b$ with $\beta_0$ the rate constant at a selected sample temperature $T_0$. After substituting $\beta$ in Eq. 3 with this function and $h_0$ with $-b/6$, the number and temperature evolutions can be fitted again to obtain the $\beta_0$ values. For the data set of non-reactive molecules in Fig. 3, this kind of modeling gives $\beta_0 = 3.4(2) \times 10^{-10}$ cm$^3$s$^{-1}$ at $T_0 = 0.97(9)$ µK. For the reactive molecules, $\beta_0 = 2.7(2) \times 10^{-10}$ cm$^3$s$^{-1}$ at $T_0 = 0.97(2)$ µK are obtained. Within the mutual error bars, the two $\beta_0$ values obtained this way agree with the corresponding data points around 0.97 µK in Fig. 4. This agreement verifies self-consistently that the $\beta$ values obtained by the segment fitting method are reasonable.

**Comparison with close-coupling quantum calculations**
For the reactive case, the loss can be well understood with the universal model in which a unity reaction probability in the short range is used, thus the rate constant and its temperature dependence are determined by the long-range collision dynamics (*3*, *30–33*). However, understanding the unexpected loss of the non-reactive molecules is more challenging. So far, the complex-mediated collision model is the only one attempting to explain it (*12*). For collisions (reactive or non-reactive alike) that proceed over a potential well, it is well known that a collision complex can be formed along the path (*1*). In the current experiment, due to the lack of convenient detection methods for the complex, its formation will manifest as loss of the NaRb molecules. To model this, following the conclusions from the highly-resonant scattering picture for non-reactive molecules (*12*), we choose a full probability into the formation of the complex at the short range. In Ref. (*12*), it was shown that this is the same short-range boundary condition used in the universal model for reactive collisions, i.e., complex formation has the same contribution to the loss as chemical reaction; thus, the loss rate constant from complex formation is also dictated by the long-range intermolecular interaction.

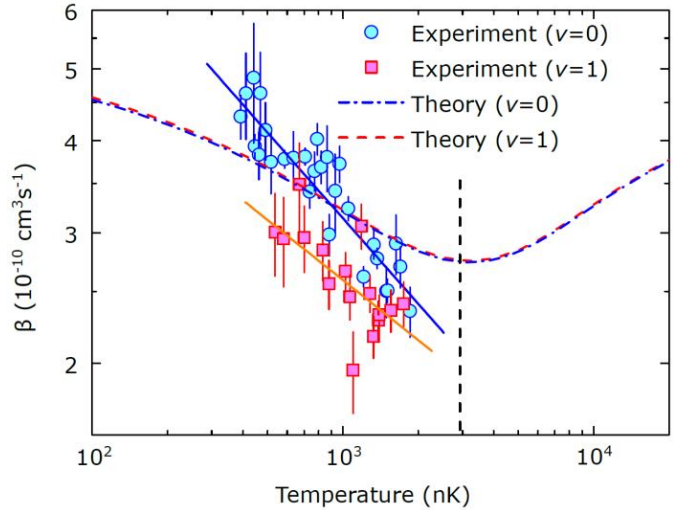

**Fig. 4. Temperature dependence of $\beta$ for different chemical reactivities.** Each $\beta$ is obtained from a fit to Eq. 3 to a segment of one full loss and heating measurement. The solid lines are from fits of $\beta$ to power law functions of $T$. Theoretical results based on the CC calculation are also shown. The dotted vertical line marks the position of $T_{\text{vdW}}$. The error bars represent one standard deviation.

Based on this model, we performed close-coupling (CC) quantum calculations for both ($v = 0$, $J = 0$) and ($v = 1$, $J = 0$) NaRb molecules using a time-independent quantum formalism, including the rotational structure of the molecules and a partial wave expansion (*34*). As shown in Fig. 4, in terms of the absolute value of $\beta$, semi-quantitative agreements between the calculations and the data can already be obtained without any free parameters. However, from the calculation, since the $C_6$ coefficient between two ($v = 0$, $J = 0$) NaRb molecules is nearly identical to that between two ($v = 1$, $J = 0$) NaRb molecules (*22–24*), the complex formation rate constant for non-reactive molecules (blue dashed curve) is essentially the same as the loss rate constant for reactive molecules (red dotted curve). This actually cannot explain the apparently larger $\beta$ measured for non-reactive molecules. This discrepancy could be reflecting that the short-range physics for chemical reaction and complex formation are not exactly identical. Although we have not attempted in the current work, it is indeed possible to modify the rate constants for better agreements if more elaborate short-range model, including a non-unity reaction probability and the phase of the reflected scattering wave, were considered (*34*).

**Possibility of post-complex-formation dynamics**
Another possible reason for the faster loss of ($v = 0$, $J = 0$) NaRb molecules is the complex-molecule collision which is actually proposed as the main cause for the loss of nonreactive UPMs in Ref. (*12*) as a result of the long complex lifetime $\tau$ for non-reactive UPMs. In addition, the complex can also dissociate at a rate of $1/\tau$ to cause a revival of the original molecules. Intuitively, if the complex-molecule collision induced loss is faster than the revival from complex dissociation, it will cause additional loss of non-reactive molecules besides the complex formation and results in a larger $\beta$ following the two-body fit.

That the complex lifetime could be very long is a unique feature for non-reactive UPMs. This is because the total angular momentum for a pair of colliding non-reactive ultracold absolute ground-state molecules, with both the internal rotation $J = 0$ and the external relative motion $l = 0$, is zero. Because of energy and angular momentum conservations, the number of open scattering channels for the complex, $N_o$, can only be one; in other words, the complex can only exit the potential well by dissociating back to the original molecules. On the other hand, near the collision threshold, we estimated using the same procedure as in Ref. (*12*) that the density of states $\rho$ of the rotational and vibrational motion of the four-atom complex $Na_2Rb_2^*$ is on the order of $10^3/\mu K$ to $10^4/\mu K$, which is very large. Thus, following the Rice-Ramsperger-Kassel-Marcus (RRKM) theory (*1*), the lifetime of the complex $\tau = 2\pi\hbar\rho/N_o$ could be 0.1 second to several seconds long. We emphasize that long complex lifetime is only possible for non-reactive UPMs. For the chemical reactive case, $\tau$ should still be short even at ultracold temperatures due to the large $N_o$ from the many possible vibrational and rotational $Na_2$ and $Rb_2$ product levels. This short complex lifetime excludes both the complex-molecule collision and the complex revival, thus the two-body model (Eq. 3) is enough to describe the reactive loss.

Confirming the complex-mediated collisions rigorously requires a detailed comparison between the experimental data and the full complex-mediated collision model. This model consists of two coupled rate equations, one for the molecule including losses from the complex formation and complex-molecule collision, and the revival; the other one for the complex including the complex formation and losses from both the complex-molecule collision and the complex dissociation (*12*). Unfortunately, due to the heating, the variation of $\beta$ with temperature, and especially the unknown distribution of the complex and the possibly temperature-dependent complex-molecule rate constant, comparing our measurement to this model is very involved. Assuming the complex and molecule always have the same temperature, and the complex has twice the polarizability of the molecule, but neglecting the temperature dependence of both rate constants, we constructed a model with the heating taken into account. The best fit of the non-reactive loss and heating data in Fig. 3 with this model results in a complex lifetime of $\tau = 0.038(6)$ s, which is shorter than the lower bound of our estimation from the RRKM theory, and a complex-molecule loss rate constant of $4.4(6) \times 10^{-9}$ cm$^3$s$^{-1}$, which is more than one order of magnitude larger than the s-wave unitary limit and is thus non-physical. Although this seems to support the absence of the complex-mediated collisions, we think it is still not conclusive due to the crudeness of the approximations made in the model.

**DISCUSSION**
The ultimate way of solving all the aforementioned difficulties is by detecting the complex and studying the post-complex-formation dynamics directly (*35*), e.g., with the very sensitive state resolved ionization detection method. Important information can also be learned by comparing the differences of collisions between two molecules and three molecules in the molecule assembly experiments with optical tweezers (*36*). Especially, with two molecules only, the complex-molecule collision is removed; thus, the complex lifetime can be directly measured by watching the time evolution of the molecule revival signal. This should allow a direct verification of the RRKM theory at the ultracold regime. Finally, besides its importance in ultracold chemistry, the existence of the complex should also be taken seriously in investigating dipolar many-body physics in optical lattices for example by considering multi-channel Hubbard models (*37*).

## MATERIALS AND METHODS

The optical trap potential is provided by a single frequency high-power fiber laser operating at 1064.4 nm. This wavelength is far below all possible excited states for molecules in the lowest two vibrational levels of the $X^1\Sigma^+$ state and is also away from all resonances for FMs. The trap frequencies $\omega_x$, $\omega_y$, and $\omega_z$ of the ground-state molecules are obtained by measuring their center-of-mass oscillations in the optical potential. The measured values agree well with calculations based on theoretical polarizabilities at 1064.4 nm, which are 674.17 a.u. for ($v = 0$, $J = 0$) and 690.16 a.u. for ($v = 1$, $J = 0$) NaRb molecules (*38*), respectively.

The ($v = 1$, $J = 0$) level lies 3172.8 GHz above the rovibrational ground state. Thanks to the fortuitously favorable transition strength, the same excited intermediate level for making ($v = 0$, $J = 0$) NaRb molecules with STIRAP previously (*11*) can still be used here; thus, we only need to tune the dump laser frequency for creating ($v = 1$, $J = 0$) molecules.

For the current investigation, it is crucial to prepare molecules in a single quantum state. Due to the atomic nuclear spins ($I_{Na} = I_{Rb} = 3/2$), for the $J = 0$ states, there are 16 hyperfine Zeeman levels distributed in a small frequency span of ∼1 MHz. As shown in Fig. 2A and B, these hyperfine structures, especially the lowest energy ones for ($v = 0$, $J = 0$) and ($v = 1$, $J = 0$) with $M_F = 3$ used in this work (marked by open circles), can be fully resolved with carefully chosen Raman laser power and polarization combinations. Here $M_F = m_I^{Na} + m_I^{Rb}$ is the projection of the total angular momentum, with $m_I^{Na}$ and $m_I^{Rb}$ the projections of the Na and Rb nuclear spins. Experimentally, up to $1.5 \times 10^4$ NaRb molecules in either $M_F = 3$ hyperfine levels can be created routinely with STIRAP efficiencies around 93% and number fluctuations of typically less than ±5%. We have verified experimentally that the other about 7% population not in the ground state has negligible effects.

**Acknowledgments**

**General**: We are grateful to Olivier Dulieu and Roman Vexiau for the valuable discussions and the polarizability calculation.

**Funding:** This work is supported by the COPOMOL project which is jointly funded by Hong Kong RGC (grant no. A-CUHK403/13) and France ANR (grant no.ANR-13-IS04-0004-01). The Hong Kong team is also supported by the National Basic Research Program of China (grant No. 2014CB921403) and the RGC General Research Fund (grant no. CUHK14301815).

**Author contributions:** X.Y., M.G. and D.W. carried out the experiment and analyzed the data. M.L.G.M and G.Q performed the close-coupling calculation. X.Y. and D.W. wrote the manuscript with inputs from all authors.

**Competing interests:** The authors declare that they have no competing interests.

**Data and materials availability:** All data needed to evaluate he conclusions in the paper are present in the paper and/or the Supplementary Materials. Additional data related to this paper may be requested from the authors at djwang@cuhk.edu.hk.

# Supplementary Materials for

## Collisions of Ultracold $^{23}$Na$^{87}$Rb Molecules with Controlled Chemical Reactivities

Xin Ye, Mingyang Guo, Maykel L. González-Martínez, Goulven Quéméner, Dajun Wang∗

**This file includes:**

- Section S1. Extracting β with segment fit
- Section S2. Dependence of β on STIRAP efficiency
- Section S3. The complex-mediated collision model
- Figs. S1 to S3

## Section S1. Extracting $\beta$ with segment fit

As shown in the example in Fig. S1, to extract the $T$ dependence of $\beta$, we first pick four segments from a whole measurement based on the criteria that the change of $T$ is less than 20% during each segment. We assume the change of $\beta$ is small in each of these segments. Only data from the first 100 ms are used since we cannot measure the temperature reliably beyond that.

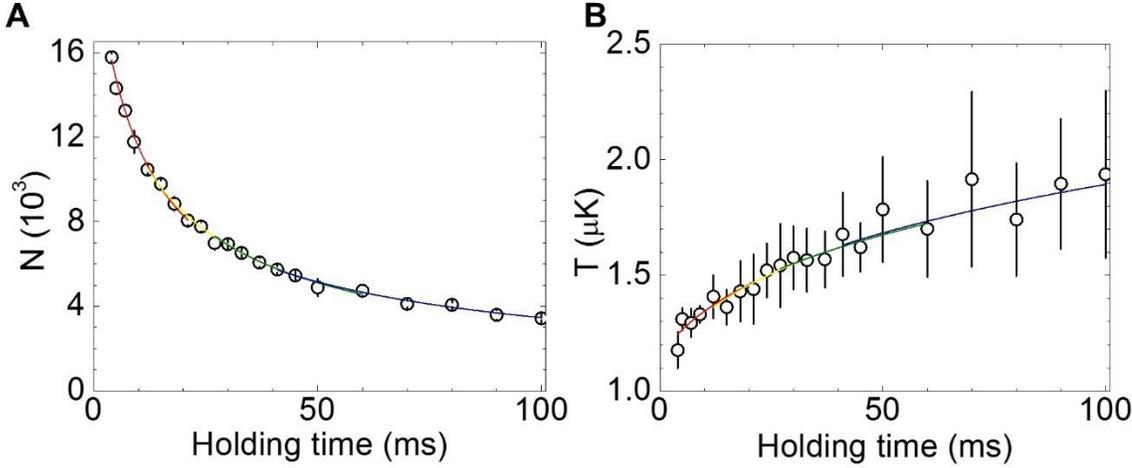

**Fig. S1. Segment two-body fit for extracting the temperature dependence of $\beta$.** (A) Number evolution and (B) temperature evolution data with curves obtained from the segment fit. In this example, the red, yellow, green and blue curves cover segments from 4 ms to 21 ms, 12 ms to 33 ms, 27 ms to 60 ms, and 41 ms to 100 ms, respectively.

For each segment, the evolutions of $N$ and $T$ are fitted simultaneously with the two-body loss model (Eq. 3) by minimizing the total dimensionless $\chi^2$, defined as

$$\chi^2 \equiv \sum_{y=N,T;i} \frac{(y_i - \bar{y}_i)^2}{\Delta y_i^2}. \tag{S1}$$

Here $y_i$ is the fitting value, $\bar{y}_i$ and $\Delta y_i$ are the mean and the standard deviation of the measurements respectively. The free parameters of the fit are the initial number, the initial temperature, $\beta$ and $h_0$. In Fig. S1, the fitting results of each segments are color coded.

## Section S2. Dependence of $\beta$ on STIRAP efficiency

Since the STIRAP efficiency is never 100%, there is always a small portion of the FMs that leaves the Feshbach state but does not reach the target state. These molecules may occupy a series of rovibrational levels of the $X^1\Sigma^+$ and $a^3\Sigma^+$ potentials and cause extra losses for the ground-state molecules via inelastic collisions. To evaluate this effect, we measure $\beta$ of the ground-state molecules at STIRAP efficiencies from 63% to 91%. The STIRAP efficiency is controlled by adjusting the Raman laser power. As can be seen from Fig. S2, $\beta$ only has a very weak dependence on the STIRAP efficiency. A linear fit gives an increase of 1.8% in $\beta$ per 10% decrease of STIRAP efficiency. Thus, at a typical STIRAP efficiency of more than 90%, the influence of the molecules not in the target level can be neglected.

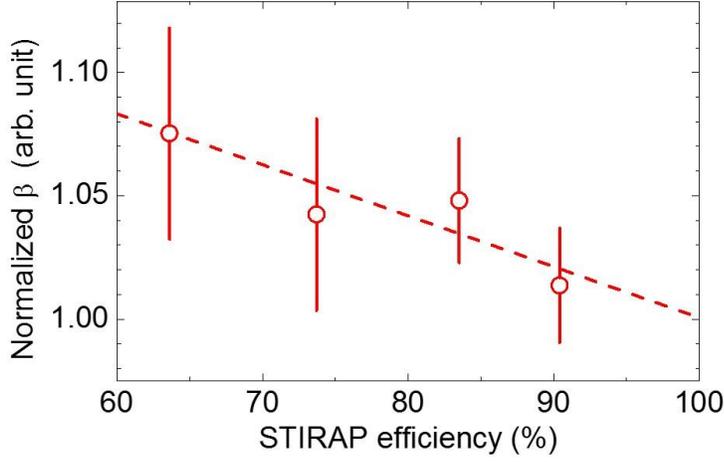

**Fig. S2. Dependence of $\beta$ on STIRAP efficiency.** As shown are the normalized $\beta$ at different STIRAP efficiencies for molecules in the absolute ground state. The dashed line is a linear fit which gives a 1.8% increment in $\beta$ per 10% decrease of STIRAP efficiency.

**Section S3. The complex-mediated collision model**

The complex-mediated collision model in ref. [12] is described by two coupled rate equations of the densities of the molecule $n$ and the four-atom complex $n_c$,

$$\frac{dn}{dt} = -\beta_{mm}n^2 + 2\gamma_c n_c - \beta_{cm}n_c n \tag{S2}$$

$$\frac{dn_c}{dt} = \frac{1}{2}\beta_{mm}n^2 - \gamma_c n_c - \beta_{cm}n_c n, \tag{S3}$$

with $\beta_{mm}$ and $\beta_{cm}$ the complex formation and complex-molecule collision rate constants respectively, and $\gamma_c = 1/\tau$ the complex dissociation rate. In this model, both the two-body complex formation and the complex-molecule collision cause loss of molecules, while the complex dissociation shows as a revival to the molecule signal.

To take the heating into account, we derived an extended model starting from the above rate equations. Since we cannot detect the complex, the following assumptions were made in the derivation: 1) the polarizability of the complex is twice that of the molecule, thus the trap frequency of the complex is the same as that of the molecule; 2) the complexes are always thermalized with the molecules; 3) both $\beta_{mm}$ and $\beta_{cm}$ are temperature independent. We note that the extended model just represents the best of our effort since these assumptions are not guaranteed to hold. Our complex-mediated collision model with heating consists of three coupled rate equations for the molecule number $N$, the complex number $N_c$, and the temperature $T$,

$$\frac{dN}{dt} = -A_{mm}\beta_{mm}\frac{N^2}{T^{3/2}} + 2\gamma_c N_c - A_{cm}\beta_{cm}\frac{N_c N}{T^{3/2}} \tag{S4}$$

$$\frac{dN_c}{dt} = \frac{1}{2}A_{mm}\beta_{mm}\frac{N^2}{T^{3/2}} - \gamma_c N_c - A_{cm}\beta_{cm}\frac{N_c N}{T^{3/2}} \tag{S5}$$

$$\frac{dT}{dt} = \frac{\frac{1}{4}A_{mm}\beta_{mm}N^2 + \frac{1}{2}A_{cm}\beta_{cm}N_cN}{(N+N_c)T^{1/2}} - \frac{\frac{1}{2}\gamma_c N_c T}{N+N_c}, \tag{S6}$$

with $A_{mm} = (\bar{\omega}^2 m/4\pi k_B)^{3/2}$, $A_{cm} = (\bar{\omega}^2 m/3\pi k_B)^{3/2}$. Fig. S3 shows the fit of the $(v=0, J=0)$ data in the Fig. 3 of the main text with the above model.

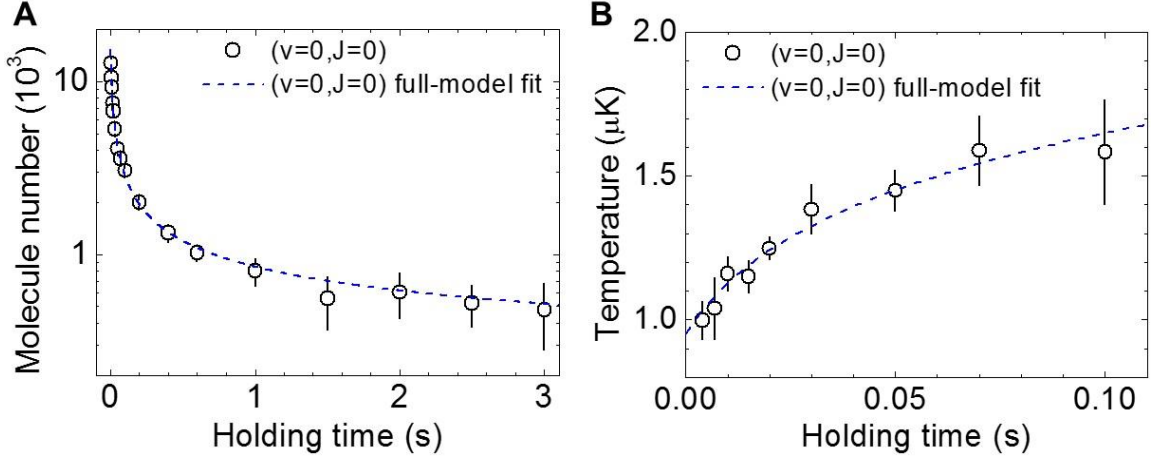

**Fig. S3. Checking the complex-mediated collisions.** The same ($v = 0$, $J = 0$) data as in the Fig.3 of the main text are fitted with Eqs. (S4), (S5) and (S6). We obtain $\beta_{cm} = 4.4(6) \times 10^{-9}$ cm$^3$s$^{-1}$ and $\tau = 0.038(6)$ s from the fit.